\documentclass[12pt]{article}
\usepackage{amsbsy,amssymb}

\usepackage[english,ukrainian]{babel}

\newcommand{\be}{\begin{eqnarray}}
\newcommand{\ee}{\end{eqnarray}}
\newcommand{\ben}{\begin{eqnarray*}}
\newcommand{\een}{\end{eqnarray*}}
\newcommand{\la}{\langle}
\newcommand{\ra}{\rangle}

\title{
A SELF-CONSISTENT  THEORY OF LIQUID $^4$He}
\author{I. O. Vakarchuk\\
{\small Ivan Franko National University of Lviv}\\
{\small 12 Drahomanov Street, Lviv UA-79005, Ukraine}\\
{\small E-mail: chair@ktf.franko.lviv.ua}}
\date{ }
\begin{document}
\maketitle

\begin{abstract}
A new method is proposed for the calculation of full density
matrix and thermodynamic functions of a many-boson system.
Explicit expressions are obtained in the pair correlations
approximation for an arbitrary temperature.
The theory is self-consistent in the sence that the calculated
properties at low temperatures coincide with that of
Bogoliubov theory and in the high-temperature limit lead to the
results for classical non-ideal gas in the random phase
approximation. The phase transition is also revealed as a
concequence of Bose--Einstein condensation deformed by interatomic
interactions.
All the final formulae are written solely via the liquid structure factor
taken as a source information instead of the interatomic potential
and, therefore, interconnect only observable quantities. This
gives also a possibility to study such a strongly non-ideal system
as liquid $^4$He.
\end{abstract}

\vspace{0.5cm}


\vspace{0.5cm}

\newpage

\section*{INTRODUCTION}

    The properties of liquid helium are known to have brought about
ample literature. Yet due to its unique characteristics this quantum fluid
keeps attracting the attention of specialists in the field of theoretical and
experimental physics alike.
    The first principle microscopic description of thermodynamic and
structure functions as well as the phenomenon of Bose--Einstein
condensation at a considerable distance from the absolute zero, in
particular in the vicinity of the $\lambda$-transition, is still a problem which
cannot be taken as solved to the very end. The latter statement can be
supported by a fairly instructive  study of heat capacity in the vicinity of
the $\lambda$-transition. The $\lambda$-like form of the heat capacity in the vicinity of
the phase transition of liquid helium into the superfluid state has been
taken for  a logarithmic  ddivergence with the critical exponent $\alpha \to 0$.
This view found its way into both text-books and monographs \cite{Huang,Br,March,Is,St,Feynm}.
Precise experiments helped to find out that in fact there is no divergence
in the heat capacity \cite{Lipa96,Lipa00} even though the exponent $\alpha$ is indeed a small
but negative number, $\alpha=-0.01056$ \cite{Lipa00}.  The studies of the $\lambda$-transition based
on the renormalization group method \cite{Wils} provide us with a possibility to
carry out a correct calculus for the so-called universal characteristics
solely, i.\,e., the critical characteristics of the thermodynamic functions
and the relations of the amplitudes of their leading asymptotics at the
temperature on either side tending to the phase transition point. Even
though the thermodynamic potential functional from the two-component
order parameter for liquid  $^4$He was calculated precisely owing to the
coherent states depiction \cite{Lang,Vak78}, yet its subsequent simplifications
necessary for the implementation of the renormalized  group approach
make it impossible to describe the system's characteristics outside the
closest vicinity of the phase transition point using the same method.
Notwithstanding tangible efforts of the researchers the renormalization
group method did not yield the logarithmic divergence  of the heat
capacity (the $\alpha$ exponent was received as a small but still finite positive
number, the power divergence having been obtained). Only in the
subsequent studies which made use of the summation procedure of the
Borel perturbation theory divergent series established the negative value
of the exponent: $\alpha=-0.01294$ \cite{Kl99}, $\alpha=-0.0150$ \cite{Cam00}, $\alpha=-0.01126$ \cite{Kl00}.

    The fact that  heat capacity at the $\lambda$-point acquires a finite value
finds itself in agreement with F.~London \cite{London} in the sense that the
$\lambda$-transition in liquid $^4$He is a Bose-condensation  deformed by the
interatomic interaction inherent of the ideal Bose-gas bringing about the
steepness of the heat capacity curve in the vicinity of the
Bose-condensation  point. It is not easy to reveal the connection between
superfluidity and Bose-condensation unequivocally. This connection is a
lot more complex than a simple correspondence. For instance, at the
temperature equaling absolute zero all the atoms of the ideal Bose-gas
find themselves in the states with a zero momentum thus forming the
one-hundred per cent Bose-condensate (BC). However, such a system is
not superfluid. Conversely, liquid $^4$He at the absolute temperature zero is
superfluid even though the number of atoms in the BC, as  showed both
by theoretical and experimental studies, is but a small part of the entire
number of atoms. It is also well-known that one- and two-dimensional
systems tend to reveal superfluid characteristics with the BC being
absent. Here we are concerned with the rise of the so-called non-diagonal
long-range order when the one-particle density matrix in the coordinates
description dwindles in compliance with the power law rather than the
exponent law, at the matrix arguments being disentangled over
infinity. On the other hand a mere assumption that the BC exists in the
weakly imperfect Bose-gas made it possible for Bogoliubov \cite{Bogol} to
obtain from the first principles the energy spectrum whose properties are
close to those of the liquid $^4$He.

    One other fairly complex task lies in according Bogoliubov's
theory \cite{Bogol}  well-suited for the low temperatures  with the random phase
approximation in the theory of classical systems. It may be possible to
find expressions for the thermodynamic and structure functions of the
Bose-liquid  which would yield the results of Bogoliubov's theory for the
temperatures $T\to 0$~K. For high temperatures in the quasi-classical $\hbar \to 0$
limit this expression would bring closer to the random  phase
approximation of the classical non-ideal gas theory. It might be supposed
that such an expression would also give good results in the intermediate
temperature area where the transition point  of  liquid $^4$He into
the superfluid state is located.
    It is clear that we are referring not to the general formulae at the
level of definitions which just cannot ``be brought to  a certain
number", but to the discovery of such a first principles method of calculating the
thermodynamic properties of the many-boson system which starts from
the $N$ particles Hamiltonian and makes it possible to put forward a
regular perturbation theory with a consistent consideration of
many-particle correlations starting with the two-particle correlations.

    In the present paper we mean to dwell upon one  of the
possibilities of solving these problems. Some of the results to be
presented here were briefly summarized in \cite{Vak97} just as an
illustration of certain tricks applicable to the wave functions and the
statistical operator \cite{Vakp}. However, the full scope of these findings has
never been published. We shall calculate explicitly the complete density
matrix taking into account the two-particletwo-particle  interatomic correlations as well as
the Helmholz free energy, the energy and also
a pair liquid structure factor which will abide by the  above requirements.

\section{INITIAL EQUATIONS}
\setcounter{equation}{0}

    Let us consider the $N$ set of spinless Bose-particles with the mass
$m$  with the Hamiltonian
\be
\hat H=\sum_{j=1}^N {\hat {\bf p}_j^2\over 2m}
+\sum_{1\le i< j\le N} \Phi (|{\bf r}_i-{\bf r}_j|),
\ee
where the first term is the operator of the kinetic energy, $\hat{\bf
p}_j$ is the operator of the momentum of the $j$-th particle. The second term presents
the potential energy of the two-particle interaction between the particles with
the coordinates $x=({\bf r}_1,...,{\bf r}_N)$.  The particles movement is limited within a
 volume of $V$.

Let $\psi_n(x)$ be taken for a system of eigenfunctions of the Hamiltonian $\hat H$
and $E_n$ are its eigenvalues. Let us consider such an equation for the
statistical operator:
\be
e^{-\beta \hat H}\psi_n(x)=e^{-\beta E_n}\psi_n (x),
\ee
where $\beta=1/T$, $T$ is the temperature of the considered system of
particles. Let us introduce a certain arbitrary function $\varphi=\varphi(x)$ and let us multiply
the left-hand side of equation (1.2) by $\varphi(x)$. We will integrate it over all
coordinates $x$:
\be
\int \varphi(x)e^{-\beta \hat H} \psi_n(x) dx=e^{-\beta E_n} \int
\varphi(x)\psi_n(x)dx,
\ee
where
\ben
\int dx=\int d{\bf r}_1\ldots \int d{\bf r}_N.
\een
We will impose the following condition for the function $\varphi(x)$:
\be
\int \varphi(x)\psi_n(x)dx\neq 0.
\ee
Then making use of the self-conjugation of the operator $\hat H$ we shall
transfer it in equation (1.3) from the function $\psi_n(x)$ to $\varphi(x)$ multiplying the
l.\,h.\,s. side of this equation by $\psi_n^{*}(x')$  and summing it by all the indices of the
$n$ states:
\ben
\sum_n\int \psi_n^{*}(x')\psi_n(x) e^{-\beta \hat H} \varphi(x)dx
=\int \varphi(x)\sum_n \psi_n^{*}(x')e^{-\beta E_n} \psi_n(x)dx.
\een
We make use of the completeness condition
$$\sum_n\psi_n^{*}(x')\psi_n(x)=\delta(x'-x),$$
and also proceeding from the previous equation we will obtain
the following equation with the permutation  of the variables $x$ for $x'$ and
inversely:
\be
e^{-\beta \hat H} \varphi(x)=\int \varphi(x')R_N(x'|x)dx',
\ee
where
\be
R_N(x'|x)=\sum_n \psi_n^{*}(x) e^{-\beta E_n} \psi_n(x')
\ee
 is the density matrix in the coordinate representation.

In true fact  equation (1.5) is self-evident and could have been written
out at once as a matrix presentation of the statistical operator action over
any function $\varphi(x)$. In the Dirac notations this equation can be written
as follows
\be
\la x|e^{-\beta \hat H}\varphi\ra&=&\int \la x| e^{-\beta \hat H}|
x'\ra \la x'|\varphi\ra dx',\nonumber\\
\\
\la x|e^{-\beta \hat H}|x'\ra&=&R_N(x'|x), \ \ \la x'|\varphi\ra=\phi(x').
\nonumber
\ee

It is clear that formula (1.5) or (1.7) is applicable both to one-particle and
two-particles problems.  Notwithstanding the fact that it is obvious, this
formula makes it possible to fully restore the density matrix  (1.6). Here
most significant heuristically is an arbitrary choice of the function $\varphi(x)$ on
condition that it is non-orthogonal to the eigen-functions of the
$\hat H$ operator and the existence of the integral (1.4).

The present article is concerned with the calculus of the $R_N(x'|x)$
density matrix for the studied many-boson system from the initial
equation (1.5).

    Should the interparticle interaction be switched off, the
density matrix will transfer into that of the ideal Bose-particles $N$ system.
That is why we will try to construct the density matrix of the $N$ system of
interacting particles as a product of the density matrix of the ideal
Bose-gas $R_N^0(x'|x)$ and the $P_N(x'|x)$ factor taking into account the interparticle
interaction:
\be
R_N(x'|x)=R_N^0(x'|x)P_N(x'|x),
\ee
where at temperature $T\neq 0$
\be
R_N^0(x'|x)={1\over N!} \left({m^{*}\over 2\pi \beta
\hbar^2}\right)^{3N/2}\sum_Q \exp \left[-{m^{*}\over 2\beta \hbar^2}
\sum_{j=1}^N ({\bf r}'_j-{\bf r}_{Qj})^2\right],
\ee
the summation over $Q$ is at the same time the summation over all the
permutation $N!$ numbering the  particles coordinates; at $T=0$, when all the particles
have zero momenta, the matrix $R_N^0(x|x')=1/V^N$. The mass of the
particle in (1.9) is understood as a certain effective mass $m^{*}$ which
equals the initial mass $m$ unless the interaction  takes place.
Thus a part
of the contribution from the interparticle interactions is taken into
account by renormalizing the particle mass, the remainder being left in the
$P_N(x'|x)$ factor. Postulating this for the density matrix (1.8) is justified also
by the fact that it was obtained under certain assumptions by a different
method in \cite{JPS97} by a direct calculus from (1.6). The expression for the
effective mass was found to be in agreement with the formula for the
effective mass of the impurity atom moving in the Bose-liquid if the
initial impurity mass coincides with the mass of the atom of liquid.
Yet for the $R_N(x'|x)$ matrix in  \cite{JPS97} an expression was found  which is correct
only at $T\to 0$~K. Here we will try to find a formula for $P_N(x'|x)$ which
will be capable of working in the entire temperature interval. We will
also postpone the discussion of the issue of determining the effective
mass contained in expression (1.9).

    Falling back on the phenomenological consideration concerning
the insignificance of the atoms permutation which are located at the
distances smaller than their own sizes quite some time ago R.~Feynman
\cite{Feynman53,Feynm} constructed the $N$-particle distribution function of the liquid $^4$He,
i.\,e., the diagonal elements of the density matrix as a product  of the ideal
Bose-gas distribution function multiplied by the factor accounting for the
atom impenetrability. We would like to base our analysis on the precise
equation (1.5) and suggest a consistent method for the calculus of the $P_N(x'|x)$
function computing the latter explicitly accounting for the two-particle
correlations.
    We will choose the $P_N(x'|x)$ matrix which takes into account the interparticle correlations as follows:
\be
P_N(x'|x)=\exp \Bigg\{c_0+\sum_{{\bf q}\neq 0}c_1(q) \rho'_{\bf q} \rho_{-{\bf q}}
-{1\over 2} \sum_{{\bf q}\neq 0} c_2(q) \Bigl[\rho_{\bf q}\rho_{-{\bf
q}}+\rho'_{\bf q}\rho'_{-{\bf q}}\Bigr]
\Bigg\},
\ee
where the Fourier coefficients of the particles density fluctuations
\be
\rho_{\bf q}={1\over \sqrt N} \sum_{j=1}^N e^{-i{\bf qr}_j},
\ee
\be
\rho'_{\bf q}={1\over \sqrt N} \sum_{j=1}^N e^{-i{\bf q}{\bf r}'_j},
\ee
at ${\bf q}\neq 0$. The components of the wave vector ${\bf q}$ cover the integer values
devisible by $2\pi/V^{1/D}$  where $D$ is dimentionality of the cubic box which
contains the system of particles studied. We will determine the
coefficient functions $c_0$, $c_1(q)$, $c_2(q)$ from equation (1.5).

We have confined ourselves to the consideration of two-particle
interparticle correlations in expression (1.10). Accounting of the
three-particle and higher correlations is accomplishable by adding the members
with a product of three or more $\rho_{\bf q}$ in the exponent  in (1.10).
In our work we  will not take them into account explicitly, yet we will
return to them when  discussing the issue of the effective mass $m^{*}$. The
exponent form of the $P_N(x'|x)$ matrix is caused by the classical boundary of its
diagonal elements when $P_N(x|x)$ turns into the usual Boltzman factor
$\exp(-\beta \Phi)$, where $\Phi$ is the potential energy of the interparticle interaction thus
equaling  the second term in the Hamiltonian (1.1).

Finally, we must also choose the appropriate $\varphi$ function which is
contained  in equation (1.5) and meets requirement (1.4). In our case we
will adjust it from the class of such functions:
\be
\varphi(x)=\exp \left[{\sum_{{\bf q}\neq 0}\lambda(q)\rho_{\bf q}}\right],
\ee
where the arbitrary coefficient function $\lambda(q)$ is a real function from the
wave vector module $q=|{\bf q}|$ with all the necessary properties.

\section{THE LEFT-HAND SIDE OF EQUATION (1.5)}
\setcounter{equation}{0}

Let us represent the action of the  statistical operator on the $\varphi$ function
in (1.5) in the form of the exponent function:
\be
e^U=e^{-\beta \hat H}\varphi.
\ee
Here again when fixing the exponent form of this function we proceed
from the assumption that in the quasi-classical limit when $\hbar\to 0$ the
statistical operator comes up to the Boltzman factor which just as the $\varphi$
function has an exponent form. Differentiating either sides of the
equation with respect to inverse temperature $\beta$ we will find the equation for the
unknown function $U$:
\ben
-{\partial U\over \partial \beta}=e^{-U} \hat H e^U,
\een
or explicitly
\be
-{\partial U\over \partial \beta}=-{\hbar^2\over 2m} \sum_{j=1}^N
\left[\mbox{\boldmath$\nabla$}_j^2U+(\mbox{\boldmath$\nabla$}_j
U)^2\right]+\Phi,
\ee
where from (2.1) and (1.13) it is apparent that
\be
U=\sum_{{\bf q}\neq 0} \lambda (q)\rho_{\bf q},\ {\rm when}\ \ \ \beta=0.
\ee

To solve equation (2.2) we will choose the function $U$ in the form of the
series ``by the  degrees" of the $\rho_{\bf q}$ quantities:
\be
U=a_0+\sum_{{\bf q}\neq 0}a_1(q) \rho_{\bf q} +
{1\over 2} \sum_{{\bf q}\neq 0} a_2(q) \rho_{\bf q}\rho_{-{\bf
q}}+\ldots\ .
\ee
We have agreed to take into account the two-particle interparticle correlation
only that is why the series is truncated at the second degree of $\rho_{\bf q}$.
Higher correlations denoted as  dots in (2.4)  can be taken into account
by summing higher degrees of $\rho_{\bf q}$. It is convenient to  work with our
equation in terms of the $\rho_{\bf q}$ quantity rather than the individual
coordinates $x$.  That is why we will show the potential energy $\Phi$
through the  quantities of $\rho_{\bf q}$:
\be
\Phi={N(N-1)\over 2V} \nu_0+{N\over 2V}\sum_{{\bf q}\neq 0}
\nu_q(\rho_{\bf q}\rho_{-{\bf q}}-1),
\ee
where the Fourier coefficients of the two particles potential energy
\be
\nu_q=\int e^{-i{\bf qR}}\, \Phi(R)\, d{\bf R}.
\ee
Equation (2.2) will now look as follows
\be
&&-{\partial U\over \partial \beta}=\sum_{{\bf k}\neq 0} {\hbar^2
k^2\over 2m}
\left(\rho_{\bf k} {\partial U\over \partial \rho_{\bf k}}
-{\partial^2 U\over \partial \rho_{\bf k}\partial \rho_{-{\bf k}}}
-{\partial U\over \partial \rho_{\bf k}}{\partial U\over \partial \rho_{-{\bf k}}}
\right)\nonumber\\
&&+{1\over \sqrt N}\mathop{ \sum_{{\bf k}\neq 0}\sum_{{\bf k}'\neq 0}}
\limits_{{\bf k}+{\bf k}'\neq 0}
{\hbar ({\bf k k}') \over 2m}
\rho_{{\bf k}+{\bf k}'} \left(
{\partial^2 U\over \partial \rho_{\bf k}\partial \rho_{{\bf k}'}}
+{\partial U\over \partial \rho_{\bf k}}{\partial U\over \partial \rho_{{\bf k}'}}
\right)\nonumber
\\
&&+{N(N-1)\over 2V}\nu_0+ \sum_{{\bf k}\neq 0} {N\over 2V} \nu_k
(\rho_{\bf k}\rho_{-{\bf k}}-1).
\ee
We substitute in this equation expression (2.4) for the $U$ function and
from the equivalence condition of the coefficient functions at the equal
degrees of $\rho_{\bf k}$  in the left- and right-hand sides of this equation  and
again taking into account just the two-particle correlation we find a system of
three equations for the unknown quantities $a_0$, $a_1(q)$ and $a_2(q)$
\be
-{d a_0\over d\beta}=-\sum_{{\bf q}\neq 0} {\hbar^2 q^2\over 2m}
[a_2(q)+a_1^2(q)]+{N(N-1)\over 2V} \nu_0-\sum_{{\bf q}\neq 0}
{N\over 2V}\nu_q,
\ee
\be
&&-{d a_1(q)\over d\beta}={\hbar^2 q^2\over 2m}
a_1(q)[1-2a_2(q)],
\\
&&-{d a_2(q)\over d\beta}={\hbar^2 q^2\over m}
[a_2(q)-a_2^2(q)]+{N\over V}\nu_q.
\ee

We begin to solve the system of these equations from equation (2.10) for
$a_2(q)$. Notwithstanding the fact that this equation is nonlinear  its
solution can be found analytically   which was suggested for the first
time in \cite{Yuhn} in the framework of the so-called shifts and collective
variables method:
\be
a_2(q)=-{\alpha_q-1\over 2} {1-e^{-2\beta E(q)}\over
1+{\alpha_q-1\over \alpha_q+1} e^{-2\beta E(q)}},
\ee
\be
\alpha_q=\sqrt{1+\left.{2N\over V}\nu_q\right/{\hbar^2 q^2\over
2m}},
\ee
where
\be
E(q)=\alpha_q{\hbar^2 q^2\over 2m}
\ee
is Bogoliubov's elementary excitation spectrum \cite{Bogol}.

After this we
solve in turn equations (2.9) and (2.8). Fairly simple even though
cumbersome transformations taking into account the ``initial" condition
(2.3) lead to the following result
\be
a_1(q)=\lambda(q)
{{2\alpha_q\over \alpha_q+1} e^{-\beta E(q)} \over 1+{\alpha_q-1\over
\alpha_q+1}e^{-2\beta E(q)}},
\ee

\be
a_0&=&-\beta E_0 -{1\over 2} \sum_{{\bf q}\neq 0} {\rm ln}
\left[{1+{\alpha_q-1\over \alpha_q+1} e^{-2\beta E(q)}\over 1+{\alpha_q-1\over \alpha_q+1}}  \right]
\nonumber\\
&+&\sum_{{\bf q}\neq 0} \lambda^2(q)
{1-e^{-2\beta E(q)}\over (\alpha_q+1)\left[1+ {\alpha_q-1\over \alpha_q+1} e^{-2\beta
E(q)}\right]},
\ee
where
\be
E_0={N(N-1)\over 2V} \nu_0-\sum_{{\bf q}\neq 0} {\hbar^2 q^2\over
8m} (\alpha_q-1)^2
\ee
is the energy of the ground state  of the many-boson system  in
Bogoliubov's approximation \cite{Bogol}.

Thus in the approximation of pair correlation we have found explicitly
the result of the action of statistical operators on the $\varphi$ function, i.\,e.,
the left-hand side of our main equation   (1.5) which taking into account
(2.1) looks as follows:
\be
e^U=\int \varphi(x') R(x'|x)dx'.
\ee
To go further we must calculate the integral over $x'$ in the right-hand side
of this equation.

\section{THE RIGHT-HAND SIDE OF EQUATION (1.5)}
\setcounter{equation}{0}

Let us now proceed to the finding of the explicit form   of the right-hand
side of our main equation (2.17) making use of the postulated  form of the
(1.8)--(1.10) statistical operator. We have

\be
\int\varphi(x')R(x'|x)dx'&=&\int \!\!\ldots \int  R_N^0 (x'|x)
\exp\Biggl\{{\sum_{{\bf q}\neq 0}
\lambda(q)\rho'_{\bf q}}+ c_0+\sum_{{\bf q}\neq 0} c_1(q) \rho'_{\bf
q}\rho_{-{\bf q}}
\nonumber\\
&-&{1\over 2}\sum_{{\bf q}\neq 0} c_2(q)\Bigl[\rho_{\bf q} \rho_{-{\bf
q}}+\rho'_{\bf q} \rho'_{-{\bf q}}\Bigr]\Biggr\}d{\bf r}'_1\ldots d{\bf
r}'_N.
\ee
Let us pass over in this expression from the integration of particles ${\bf r}'_1,...,{\bf r}'_N$
by the individual coordinates to the integration over the  $\rho'_{\bf q}$ variables
which are determined by formula (1.12). Such a transition is performed
by  means of Zubarev's \cite{Zub1} transition function which is a product of the
Dirac $\delta$-function and equation (3.1) and looks as follows:
\be
&&\int\!\!\! \varphi(x')R(x'|x)dx'=\!\! \int \!\!\! \ldots \!\!\int \exp\Biggl\{ {\sum_{{\bf q}\neq 0}
\lambda(q)\rho'_{\bf q}}+ c_0+\sum_{{\bf q}\neq 0}
c_1(q) \rho'_{\bf q}\rho_{-{\bf q}}
\nonumber\\
&&-{1\over 2}\sum_{{\bf q}\neq 0} c_2(q)
[\rho_{\bf q} \rho_{-{\bf q}}+\rho'_{\bf q} \rho'_{-{\bf q}}]
\Biggr\}J_0(\rho') (d\rho'),
\ee
where the weight function
\be
J_0(\rho')=\int \ldots \int R_N^0(x'|x) \mathop{\prod_{{\bf q}\neq 0}}\nolimits' \delta
\left(\rho'_{\bf q}-{1\over \sqrt N} \sum_{j=1}^N e^{-i{\bf
qr}'_j}\right)d{\bf r}'_1\ldots d{\bf r}'_N.
\ee
An element of the $\rho_{\bf q}'$ space volume
\be
(d\rho')=\mathop{\prod_{{\bf q}\neq 0}}\nolimits' d\rho_{\bf q}'^{c} d\rho_{\bf q}'^{s},
\ee
where $\rho_{\bf q}'^{c}$, $\rho_{\bf q}'^{s}$ are  a real and respectively an imaginary part of the variable
$\rho_{\bf q}'=\rho_{\bf q}'^{c}-i\rho_{\bf q}'^{s}$.
The integration  by $\rho_{\bf q}'^{c}$, $\rho_{\bf q}'^{s}$ takes place in the infinite volume. The prime
($'$) at the product sign in (3.3) and (3.4) means that the values of the wave
vector ${\bf q}$ are considered only from the half-space of all of its possible
values because there exists the dependence $\rho_{\bf q}'^{*}=\rho_{-{\bf q}'}$
or $\rho_{\bf q}'^{c}=\rho_{-{\bf
q}}'^{c}$,\ \ \ $\rho_{\bf q}'^{s}=-\rho_{-{\bf q}}'^{s}$.

We will use the integral representation for the $\delta$-function and write the
transition function (3.3) as
\be
J_0(\rho')&=&\!\!\!\int \ldots\! \int\! R_N^0(x'|x)\!
\int (d\omega) \exp\left[\pi i\sum_{{\bf q}\neq 0} \omega_{\bf
q}\left(\rho'_{\bf q}-{1\over \sqrt N} \sum_{j=1}^N e^{-i{\bf
qr}'_j}\right)\right]
d{\bf r}'_1\ldots d{\bf r}'_N\nonumber
\\
&=&\int (d\omega) \exp\left[\pi i\sum_{{\bf q}\neq 0} \omega_{\bf
q}\rho'_{\bf q}\right] \Bigg\la \exp\left[-\pi i \sum_{{\bf q}\neq 0} {1\over \sqrt
N} \omega_{\bf q} \sum_{j=1}^N e^{-i{\bf qr}'_j}\right]\Bigg\ra,
\ee
here the element of the $\omega$ space
$$(d\omega)=\mathop{\prod_{{\bf q}\neq 0}}\nolimits' d
\omega_{\bf q}^c d \omega_{\bf q}^s,$$
where $\omega_{\bf q}^c$, $\omega_{\bf q}^s$  are a real and respectively
an imaginary part of the $\omega_{\bf q}$ complex variable
linked to $\rho_{\bf q}$, the angle brackets denoting the integration over the
individual  primed   coordinates with the weight equaling the density
matrix  of the ideal gas:
\be
\la \ldots \ra=\int \ldots \int(\ldots)
\!\!R_N^0(x'|x)d{\bf r}'_1\ldots d{\bf r}'_N.
\ee
The integral from the $R_N^0(x'|x)$ density matrix itself by $x'$ equals unity.  Indeed,
\be
&&\int\ldots\int R_N^0(x'|x)d{\bf r}'_1\ldots d{\bf r}'_N
\nonumber\\
&&=
{1\over N!} \left({m^{*}\over 2\pi \beta \hbar^2}\right)^{3N/2}
\sum_Q \int \ldots \int e^{-{m^{*}\over 2\beta \hbar^2}
\sum_{j=1}^N ({\bf r}'_j-{\bf r}_{Qj})^2}d{\bf r}'_1\ldots d{\bf
r}'_N\\ \nonumber
&&={1\over N!} \left({m^{*}\over 2\pi \beta \hbar^2}\right)^{3N/2}
\sum_Q \int \ldots \int e^{-{m^{*}\over 2\beta \hbar^2}
\sum_{j=1}^N {\bf R}_j^2}d{\bf R}_1\ldots d{\bf R}_N=1,
\ee
where we have passed to the new variables of integration
$${\bf R}_j={\bf r}'_j-{\bf r}_{Qj},$$
and at the same time we assume that the size of the box where the system
of particles is located is already infinite, $V\to \infty$. We will refer to operation
(3.6) as to averaging.

Let us now pass over to the calculation of the weight function (3.5). We
will reflect the average from the exponent in (3.5) in the form of the
exponent from the non-reducible means limiting ourselves to two-particle
correlations again:
\be
J_0(\rho')&=&\int (d\omega_{\bf q}) e^{\pi i\sum_{{\bf q}\neq 0}
\omega_{\bf q}\rho'_{\bf q}}
\exp\Biggl\{-\pi i \sum_{{\bf q}\neq 0} {\omega_{\bf q} \over
\sqrt N} \left\la \sum_{j=1}^N e^{-i{\bf qr}'_j}\right\ra
\nonumber \\
&+&{1\over 2} \sum_{{\bf q}_1\neq 0}\sum_{{\bf q}_2\neq 0}
{(\pi i \omega_{{\bf q}_1})(\pi i \omega_{{\bf q}_2})\over N}
\Bigg[ \left\la \sum_{j_1=1}^N e^{-i{\bf q}_1{\bf r}'_{j_1}}
\sum_{j_2=1}^N e^{-i{\bf q}_2{\bf r}'_{j_2}}\right\ra
\\
&-&\left\la \sum_{j_1=1}^N e^{-i{\bf q}_1{\bf r}'_{j_1}}\right\ra
\left\la\sum_{j_2=1}^N e^{-i{\bf q}_2{\bf r}'_{j_2}}\right\ra
\Bigg]+\ldots \Biggr\}.\nonumber
\ee
Let us calculate the averege quantities necessary for us. Thus similarly to (3.7) we
have
\be
&&\Big\la\sum_{j=1}^N e^{-i{\bf q}{\bf r}'_{l}}\Big\ra={1\over N!}
\left({m^{*}\over 2\pi\beta \hbar^2}\right)^{3N/2}\sum_Q
\sum_{l=1}^N \int\ldots \int d{\bf r}'_1\ldots d{\bf r}'_N
e^{-i{\bf qr}'_l} e^{-{m^{*}\over 2\beta \hbar^2} \sum_{j=1}^N ({\bf
r}'_j-{\bf r}_{Qj})^2}\nonumber\\
&&={1\over N!}\left({m^{*}\over 2\pi\beta \hbar^2}\right)^{3N/2}
\sum_Q \sum_{l=1}^N e^{-i{\bf qr}_{Ql}}
\int \ldots \int d{\bf R}_1\ldots d{\bf R}_N
e^{-i{\bf q}{\bf R}_l} e^{-{m^{*}\over 2\beta \hbar^2}
\sum_{j=1}^N R_j^2}
\nonumber\\
&&={1\over N!}\left({m^{*}\over 2\pi\beta \hbar^2}\right)^{3N/2}\sum_Q \sum_{l=1}^N
 \left({2\pi\beta \hbar^2 \over m^{*}}\right)^{3(N-1)/2}\!\!\!\!\!\!\!e^{-i{\bf q}{\bf r}_{Ql}}
\left({2\pi\beta \hbar^2 \over m^{*}}\right)^{3/2}
\\ \nonumber
&&\times e^{-\beta \hbar^2 q^2/ 2m^{*}}
=\sqrt N \rho_{\bf q}e^{-\beta \hbar^2 q^2/ 2m^{*}}.
\ee

Incidentally, it is not difficult to show that the mean for the many-fermion
system equals zero. This is connected with the fact that the $R_N^0(x'|x)$ density
matrix for the fermions  is antisymmetric as regards the permutation of the
primed $({\bf r}'_1,\ldots,{\bf r}'_N)$ coordinates, the value of (1.12) being symmetrical.
Otherwise said, for Fermi statistics the $\varphi$ function from (1.13) does not
meet  condition (1.4). Thus, it is necessary to find another function
which would be non-orthogonal to the wave functions of the many-fermion
system. The resuls of the calculation of the thermodynamic and
structure functions for the Fermi particles on the basis of the suggested
approach will be published separately.

    Our further steps are analogous to those in the case of  (3.9):
\be
&&\Big\la \sum_{j_1=1}^N e^{-i{\bf q}_1{\bf r}'_{j_1}} \sum_{j_2=1}^N
e^{-i{\bf q}_1{\bf r}'_{j_2}}\Big\ra= \Big\la \sum_{j_1=1}^N e^{-i({\bf
q}_1+{\bf q}_2){\bf r}'_{j_1}}\Big\ra +\sum_{j_1=1}^N\mathop{\sum_{j_2=1}^N}\limits_{(j_1\neq j_2)}\Big \la
e^{-i{\bf q}_1{\bf r}'_{j1}-i{\bf q}_2{\bf
r}'_{j_2}}\Big\ra\nonumber \\
&&=e^{-\beta {\hbar^2 ({\bf q}_1+{\bf
q}_2)^2\over 2m^{*}}} \sum_{j_1=1}^N e^{-i({\bf q}_1+{\bf
q}_2){\bf r}_{j1}}+ \sum_{j_1=1}^N e^{-i{\bf q}_{1}{\bf r}_{j_1}}
\sum_{j_2=1}^N e^{-i{\bf q}_{2}{\bf r}_{j_2}}
e^{-\beta\left({\hbar^2 q_1^2\over 2m^{*}}+{\hbar^2 q_2^2\over
2m^{*}}\right)}
\\
&&-e^{-\beta {\hbar^2\over 2m^{*}}
(q_1^2+q_2^2)}\sum_{j_1=1}^N e^{-i({\bf q}_1+{\bf q}_2){\bf
r}_{j_1}}. \nonumber
\ee
Now, taking into account  (3.9) and (3.10) we will find that
\be
&&\Big\la \sum_{j_1=1}^N e^{-i{\bf q}_1{\bf r}'_{j_1}} \sum_{j_2=1}^N e^{-i{\bf q}_2{\bf
r}'_{j_2}}\Big\ra-
\Big\la \sum_{j_1=1}^N e^{-i{\bf q}_{1}{\bf r}'_{j_1}}\Big\ra
\Big\la \sum_{j_2=1}^N e^{-i{\bf q}_{2}{\bf r}'_{j_2}}\Big\ra
\\ \nonumber
&&=\left(
1-e^{-2\beta{\hbar^2 q_1^2\over 2m^{*}}}
\right)N\delta ({\bf q}_1+{\bf q}_2)
+ \left[
e^{-\beta{\hbar^2\over 2m^{*}} ({\bf q}_1+{\bf q}_2)^2}
-e^{-\beta{\hbar^2\over 2m^{*}} (q_1^2+q_2^2)}\right]\sqrt N \rho_{{\bf
q}_1+{\bf q}_2}.
\ee
Here, the second term will be omitted in the accepted approximation of
two-particle correlations when substituting (3.11) in (3.8).
Expressions (3.9), (3.11) give correct asymptotics at absolute zero temperature, $\beta\to \infty$, as well.

Further, the weight function in (3.8) can be calculated easily  as the
integrals by $\omega_{\bf k}^c$  and $\omega_{\bf k}^s$ are reducible to the Poisson integrals:
\be
J_0(\rho')&=&\!\!\int (d\omega) \exp\!\left[-{1\over 2} \sum_{{\bf q}\neq 0} |\pi
\omega_q|^2 (1-e^{-2\beta{\hbar^2 q^2\over 2m^{*}}})
+\pi i \sum_{{\bf q}\neq 0} \omega_{\bf q} \left(\rho'_{\bf
q}-\rho_{\bf q}e^{-\beta{\hbar^2 q^2\over 2m^{*}}}\right)
\right]\!\nonumber\\
&=&\exp\left\{
-{1\over 2} \sum_{{\bf q}\neq 0} {\left|
\rho'_{\bf q}-\rho_{\bf q} e^{-\beta {\hbar^2 q^2\over 2m^{*}}}
\right|^2\over
1-e^{-2\beta {\hbar^2 q^2\over 2m^{*}}}}
\right\}
\mathop{\prod_{{\bf q}\neq 0}}\nolimits'{1\over \pi \left(1-e^{-2\beta {\hbar^2 q^2\over
2m^{*}}}\right)}.
\ee
Now, just as in (3.12) we can carry out integration in (3.2):
\be
&&\int \varphi(x')R(x'|x)dx'=\exp\left\{
c_0+{1\over 2}\sum_{{\bf q}\neq 0}
{\lambda^2(q)\over
\left[c_2(q)+{1\over 1-e^{-2\beta\hbar^2 q^2/
2m^{*}}}\right]}
\right\}\nonumber
\\
&&\times\exp\left\{
\sum_{{\bf q}\neq 0} \lambda(q)\rho_{\bf q} \left[
c_1(q)+
{e^{-\beta{\hbar^2 q^2\over 2m^{*}}}\over
1-e^{-2\beta{\hbar^2 q^2\over 2m^{*}}}}
\right]\left/
\left[c_2(q)+{1\over 1-e^{-2\beta{\hbar^2 q^2\over
2m^{*}}}}\right.
\right]
\right\}\nonumber\\
&&\times
\exp\Biggl\{{1\over 2}
\sum_{{\bf q}\neq 0}
{\left[c_1(q)+ {\left.e^{-\beta{\hbar^2 q^2\over 2m^{*}}}\right/
(1-e^{-2\beta{\hbar^2 q^2\over 2m^{*}}})}\right]^2
\over
\left[c_2(q)+{1\left/ (1-e^{-2\beta{\hbar^2 q^2\over
2m^{*}}})\right.}\right]}\ \rho_{\bf q}\rho_{-{\bf q}}
\nonumber\\
&&-{1\over 2} \sum_{{\bf q}\neq 0}
\left[c_2(q)+
{e^{-2\beta{\hbar^2 q^2\over 2m^{*}}}\over
1-e^{-2\beta{\hbar^2 q^2\over 2m^{*}}}}
\right]\rho_{\bf
q}\rho_{-{\bf q}}
\Biggr\}\mathop{\prod_{{\bf q}\neq 0}}\nolimits'
{1\over 1+\left(1-e^{-2\beta \hbar^2 q^2/2m^{*}}\right)c_2(q)}.
\ee
Thus, we have also found the right-hand side of the main equation (1.15).
Now we can proceed to determining the unknown coefficient functions
$c_0$, $c_1(q)$ and $c_2(q)$.

\section{EQUATIONS FOR THE
${\mbox{\lowercase{\it с}}}_{\mbox{\lowercase{\it n}}}({\mbox{\lowercase{\it q}}})$ COEFFICIENT FUNCTIONS}
\setcounter{equation}{0}

To satisfy condition (2.17) the exponent factor in (3.3) should equal the
$U$ function from (2.4) for any values of the $\rho_{\bf q}$ variable and the arbitrary
function $\lambda(q)$. From this condition we find the equation for the unknown
fucnctions $c_0$, $c_1(q)$, $c_2(q)$. Thus, we will equate the coefficients at the
identical  degrees of  $\rho_{\bf q}$ in (2.4) as well as in the exponent  of the
right-hand side part of equation (3.13):
\be
a_0&=&c_0+{1\over 2} \sum_{{\bf q}\neq 0}\lambda^2(q)\left/
\left[{ c_2(q)+{1\over 1-e^{-2\beta{\hbar^2 q^2\over
2m^{*}}}}}\right]\right.
\nonumber
\\
&-&{1\over 2} \sum_{{\bf q}\neq 0}{\rm ln} \left[1+\left(1-e^{-2\beta{\hbar^2
q^2/ 2m^{*}}} \right)c_2(q)\right],
\\
a_1(q)&=&\lambda(q)\left.
\left[c_1(q)+{e^{-\beta{\hbar^2 q^2\over 2m^{*}}}\over
1- e^{-2\beta{\hbar^2 q^2\over 2m^{*}}}}\right]\right/
\left[c_2(q)+{1\over
1- e^{-2\beta{\hbar^2 q^2\over 2m^{*}}}}\right],
\\
a_2(q)&=&\left.
\left[c_1(q)+{e^{-\beta{\hbar^2 q^2\over 2m^{*}}}\over
1- e^{-2\beta{\hbar^2 q^2\over 2m^{*}}}}\right]^2\right/
\left[c_2(q)+{1\over 1- e^{-2\beta{\hbar^2 q^2\over
2m^{*}}}}\right]
\nonumber\\
&-&\left[ c_2(q)+{e^{-2\beta{\hbar^2 q^2\over 2m^{*}}}
\over 1- e^{-2\beta{\hbar^2 q^2\over 2m^{*}}}}\right].
\ee

Let us address equation (4.1) taking into account expression (2.5) for $a_0$
as a result of the arbitrariness of the function $\lambda(q)$.  We will obtain one
more additional equation equaling the multipliers near $\lambda^2(q)$ on either side
in equation (4.1). In equation (4.2) the value of $\lambda(q)$ falls out  as can be
seen from (2.4). Thus, we now have three equations for two coefficients
$c_1(q)$ and $c_2(q)$. If our theory is consistent, one of the equations from  system
(4.1)--(4.3) will be satified identically:
\be
&&{{2\alpha_q\over \alpha_q+1}e^{-\beta E(q)}\over
1+{\alpha_q-1\over \alpha_q+1} e^{-2\beta E(q)}}={\bar c_1(q)\over
\bar c_2(q)},\nonumber\\
&&a_2(q)={\bar c_1^2(q)\over \bar c_2(q)}-\bar c_2(q)+1,\\
&&{1\over \alpha_q+1} {1-e^{-2\beta E(q)} \over 1+{\alpha_q-1\over
\alpha_q+1} e^{-2\beta E(q)}}={1\over 2 \bar c_2(q)},\nonumber
\ee
here we have introduced abbreviated notations
\be
\bar c_1(q)&=&c_1(q)+{e^{-\beta{\hbar^2 q^2\over 2 m^{*}}}\over
1-e^{-2\beta{\hbar^2 q^2\over 2m^{*}}}},\\
\bar c_2(q)&=&c_2(q)+{1\over 1-e^{-2\beta{\hbar^2 q^2\over
2m^{*}}}}.
\ee
From the third equation of system (4.4) we have
\ben
\bar c_2(q)&=&{\alpha_q+1\over 2}{1+{\alpha_q-1\over \alpha_q+1}
e^{-2\beta E(q)}\over 1-e^{-2\beta E(q)}}\nonumber\\
&=&{1+\alpha_q {\coth}[\beta E(q)]\over 2},
\een
and from (4.6)
\be
c_2(q)={1\over 2} \left\{
\alpha_q {\coth}[\beta E(q)]-{\coth} \left[\beta{\hbar^2 q^2\over 2m^*}\right]
\right\}.
\ee
Now, from the first equation  (4.4) we  find
\ben
\bar c_1(q)={\alpha_q\over 2\ \sinh[ \beta E(q)]},
\een
and from (4.5) we have
\be
c_1(q)={1\over 2}\left\{{\alpha_q\over \ \sinh[ \beta
E(q)]}- {1\over \ \sinh\left[ \beta {\hbar^2 q^2\over
2m^*}\right]} \right\}.
\ee

We can easily check now that the second equation from system (4.4) is
indeed met identically as expected.

Finally, from (4.1) we also find the coefficient
\be
c_0=-\beta E_0+{1\over 2} \sum_{{\bf q}\neq 0} {\rm ln}
\left({1-e^{-2\beta{\hbar^2 q^2\over 2m^{*}}}\over
1-e^{-2\beta E(q)}}\right)+{1\over 2} \sum_{{\bf q}\neq 0} {\rm ln}\
\alpha_q.
\ee

\section{DENSITY MATRIX AND THE PARTITION FUNCTION}
\setcounter{equation}{0}

With the help of expressions (4.7)--(4.9) taking into account (1.8)--(1.10)
we find the explicit expression for the $N$-particle density matrix:

\be
&&R_N(x'|x)=R_N^0(x'|x)\exp\!\left\{
\!\!-\beta E_0 \!+\!{1\over 2}\sum_{{\bf q}\neq 0}\! {\rm ln}\!
\left({\alpha_q\ \tanh\left[ {\beta \over 2} E(q)\right]\over
\tanh\left[ \beta{\hbar^2q^2\over
4m^{*}}\right]}\right)
+\sum_{{\bf q}\neq 0}\!{\rm ln}\!
\left({1-e^{-\beta{\hbar^2 q^2\over 2m^{*}}}\over
1-e^{-\beta E(q)}}\right)\!\!\right\}\nonumber \\
&&\times \exp\Biggl\{
-{1\over 4} \sum_{{\bf q}\neq 0}
\left(\alpha_q\ {\coth}\left[ \beta E(q)\right]-{\coth}\left[ \beta{\hbar^2 q^2\over 2m^{*}}
\right]\right)
(\rho_{\bf q} \rho_{-{\bf q}}+\rho'_{\bf q} \rho'_{-{\bf q}})
\\
\nonumber
&&+{1\over 2}\sum_{{\bf q}\neq 0} \left(
{\alpha_q\over \sinh \left[\beta E(q)\right]}-{1\over \sinh\left[ \beta
{\hbar^2 q^2\over 2m^{*}}\right]}
\right)
\rho_{\bf q} \rho'_{-{\bf q}}\Biggr\}.
\ee

By integrating the density matrix diagonal elements  in compliance with
(1.6) we obtain the partition function
\be
Z_N=\int R_N(x|x)dx,
\ee
from which we will find free energy $F=-T\ {\rm ln}\ Z_N$, and from the latter other
thermodynamic functions.

In (5.2) let us pass over from the integration over the individual
coordinates $({\bf r}_1,\ldots, {\bf r}_N)$ to the integration over the collective variables (1.11) similarly
to what was done in (3.1):
\be
Z_N&=&Z_N^0 \exp\Biggl\{ -\beta E_0+{1\over 2}\sum_{{\bf q}\neq 0}
{\rm ln} \left({\alpha\ \tanh \left[{\beta\over 2}
E(q)\right]\over \tanh \left[\beta{\hbar^2 q^2\over
4m^{*}}\right]}\right)
\nonumber\\
&+&\sum_{{\bf q}\neq 0} {\rm ln}\left(
{1-e^{-\beta{\hbar^2 q^2\over 2m^{*}}}\over 1-e^{-\beta E(q)}}
\right)\Biggr\}\int (d\rho)J(\rho)\nonumber\\ &\times&\exp \left\{
-{1\over 2} \sum_{{\bf q}\neq 0} \left(
\alpha\tanh\left[{\beta\over 2}E(q)\right]-\tanh
\left[\beta{\hbar^2 q^2\over 4m^{*}}\right] \right)\rho_{\bf
q}\rho_{-{\bf q}} \right\}, \ee \vspace*{-0.8cm}
where the weight
function
\be
J(\rho)=\Bigg\la \mathop{\prod_{{\bf q}\neq 0}}\nolimits'
\delta\Big(\rho_{\bf q}-{1\over \sqrt N} \sum_{j=1}^N e^{-i{\bf qr}_j}\Big)\Bigg\ra^0,
\ee
here the brackets marked with ``0" denote the following averaging:
\be
\la\ldots \ra^0={1\over Z_N^0} \int (\ldots) R_N^0 (x|x)dx,
\ee
and the partition function of the ideal gas
\be
Z_N^0=\int R_N^0(x|x)dx.
\ee

We calculate the $J(\rho)$ weight function from (5.4) in the same way as the
$J_0(\rho')$ function from (3.3), clearly just with a different averaging operation
(5.5) instead of (3.6). Thus, by  the respective permutation  for $J(\rho)$ we
have expressions(3.5) and (3.8). The necessary averages values contained in (3.8)
can now be calculated easily as well
\ben
\Big\la {1\over \sqrt N} \sum_{j=1}^N e^{-i{\bf qr}_j}\Big\ra^0=\sqrt
N\,
\delta_{{\bf q},0},
\een
and
\be
\Bigg\la \left({1\over \sqrt N} \sum_{j_1=1}^N e^{-i{\bf q}_1{\bf r}_{j1}}\right)
\left({1\over \sqrt N} \sum_{j_2=1}^N e^{-i{\bf q}_2{\bf r}_{j2}}\right)\Bigg\ra^0
=S_0(q_1)\delta({\bf q}_1+{\bf q}_2),\ \ {\bf q}_1\neq 0,\ {\bf q}_2\neq 0,
\ee
where $S_0(q)$ by definition is a pair structure factor of the ideal Bose-gas.
The expression for it is well-known
\be
S_0(q)=1+{1 \over N} \sum_{\bf p} n_p n_{|{\bf p}+{\bf q}|},
\ee
where
\be
n_p={1\over z_0^{-1} e^{\beta{\hbar^2 p^2\over 2m^*}}-1}
\ee
is an average number of particles  whose momentum equals $\hbar{\bf p}$, the
activity of the ideal gas $z_0$ being excluded from the condition
\be
\sum_{\bf p} n_p=N.
\ee
Let us also remember the expression for the partition function of the
ideal Bose-gas (5.6) the particles of which have the mass $m^{*}$
\be
Z_N^0=\exp\left[-\sum_{\bf q} {\rm ln} \left(1-z_0
e^{-\beta\hbar^2q^2/2m^{*}}\right)\right].
\ee

The necessary integral by $\omega_{\bf k}$ in  (3.8) taking into account pair
correlations can be easily taken. We will find  the  wave function (5.4):
\be
J(\rho)=\left(\mathop{\prod_{{\bf q}\neq 0}}\nolimits'{1\over \pi S_0(q)}\right)
\exp\left[-{1\over 2} \sum_{{\bf q}\neq 0} {\rho_{\bf q}\rho_{-{\bf q}}\over
S_0(q)}\right].
\ee

Now, taking into account (5.12) the integration over the variables $\rho_{\bf q}$ in  (5.3)
is reduced to the calculation of the Poisson integral:
\be
Z_N&=&Z_{N}^0\exp\Biggl\{
-\beta E_0+\sum_{{\bf q}\neq 0} {\rm ln}
\left({1-e^{-\beta{\hbar^2 q^2\over 2m^{*}}}\over
1-e^{-\beta E(q)}}\right)
+{1\over 2}\sum_{{\bf q}\neq 0} {\rm ln}
\left({\alpha_q \tanh \left[{\beta \over 2} E(q)\right]\over
\tanh\left[ \beta {\hbar^2q^2\over
4m^{*}}\right]}\right)\\\nonumber
&-&{1\over 2} \sum_{{\bf q}\neq 0}{\rm ln}
\left[1+\!S_0(q)\left(\alpha_q\tanh\!\!\left[ {\beta \over 2} E(q)\right]\!\!-\tanh\!\left[\beta {\hbar^2 q^2\over 4m^{*}}\right]\right)\right]\Biggr\}.
\ee

Expressions (5.1) and (5.13)  for the density matrix and for the statistical
operator are the starting formulae  for the calculation of the
thermodynamic and structure functions of the Bose-liquid.  Before we
pass over to the calculation of these values  we will make some
preliminary studies.

First of all we will say that we have not obtained the equation for the
ideal mass $m^{*}$. It is not surprising at all as we were working in the
approximation of one sum by the wave vector ${\bf q}$  and the difference
between the effective mass $m^{*}$ and the initial particle mass
$m$ is proportional to the sum by ${\bf q}$.

Let us pass to the discussion of the obtained expressions for the
partition function and the density matrix. It is apparent that when we switch
off the interparticle interaction with $\nu_q=0$, i.\,e., $\alpha_q=1$ we will obtain
from (5.1) that  for any temperature
\ben
R_N(x|x')=R_N^0(x|x'),
\een
and
$$Z_N=Z_N^0.$$
Notwithstanding the naturalness of this condition it will not be met, for
instance, by the well-known Penrose formula for the $N$-particles density
matrix. This formula was also obtained by E.~Feenberg \cite{Feen}. Applying
the method of coherent states it was also found in \cite{Lu}. With the help of
wave functions of the many-boson Bogoliubov--Zubarev system  \cite{Zub2} it
was calculated in \cite{Vak75,Vak79}. Using our notations it looks as follows:
\be
&&R_N(x|x')=\left\{{\prod_{{\bf q}\ne 0}}'\alpha_q\tanh{E(q)\over 2T}\right\}
\exp\Bigg\{-{E_0\over T}-\sum_{{\bf q}\neq 0}\ln \left(1-e^{-E(q)/T}\right)
\nonumber\\
&&+{1\over 4}\sum_{{\bf q}\neq 0} (\rho_{\bf q}\rho_{-{\bf q}}+
\rho_{\bf q}'\rho_{-{\bf q}}')
-{1\over 4}\sum_{{\bf q}\neq 0} {\alpha_q\over \sinh [E(q)/T]}
\Big[(\rho_{\bf q}\rho_{-{\bf q}}+
\rho_{\bf q}'\rho_{-{\bf q}}')
\nonumber\\
&&\times
\cosh{E(q)\over T}-(\rho_{\bf q}\rho_{-{\bf q}}'+
\rho_{\bf q}'\rho_{-{\bf q}}) \Big]\Bigg\}.
\ee
This expression holds true only for $T\to 0$. If $\alpha_q=1$, then from (5.14) we
obtain an expression for the ideal gas density matrix
\be
&&R_N^0(x|x')=\left\{{\prod_{{\bf q}\neq 0}}'\tanh\left[{\hbar^2q^2\over 4mT}\right]\right\}
\exp\Bigg\{-\!\!\sum_{{\bf q}\neq 0}\ln \left(1-e^{-\hbar^2 q^2/2mT}\right)
\nonumber\\
&&+{1\over 4}\sum_{{\bf q}\neq 0} (\rho_{\bf q}\rho_{-{\bf q}}+
\rho_{\bf q}'\rho_{-{\bf q}}')
-{1\over 4}\sum_{{\bf q}\neq 0} {1\over \sinh (\hbar^2q^2/2mT)}
\nonumber\\
&&\times
\Big[(\rho_{\bf q}\rho_{-{\bf q}}+
\rho_{\bf q}'\rho_{-{\bf q}}')\cosh\left[{\hbar^2 q^2\over 2mT}\right]-(\rho_{\bf q}\rho_{-{\bf q}}'+
\rho_{\bf q}'\rho_{-{\bf q}}) \Big]\Bigg\},
\ee
\noindent
which does not coincide with the  precise expression (1.9) for $m^{*}=m$. It is
not surprising at all as in the $\rho_{\bf q}$ representation   the kinetic energy
operator is not diagonal  (which can be shown at least from equation
(2.7)). It is just on its off-diagonal part that we are building the theory of
perturbations in which each subsequent member   of the series has in
comparison with the previous member an additional summation  by the
wave vector ${\bf q}$.

    It is curious that formulae (5.14) and (5.15) prompt to us how one
can formally obtain our result   for the density matrix. Thus we have the
equation
\be
R_N(x|x')=R_N^0(x|x'){R_N(x|x')({\rm from\ formula}\ (5.14))\over R_N^0(x|x')({\rm from\ formula}\
(5.15))}.
\ee

If we use expressions (5.14) and (5.15) in the right-hand part of this
equation for the density matrices relation, we will immediately arrive at
formula (5.1) with $m^{*}=m$. This trick also hints at a possibility of calculating
the effective mass $m^{*}$. Thus, should we find the following
approximation for the density matrix (5.14), for instance by solving
Bloch's equation directly, as suggested in \cite{Vak79}, the result will be the
following. Firstly, we will obtain in the (5.14) exponent  the
$\sim \rho_{{\bf q}_1}\rho_{{\bf q}_2}\rho_{{\bf q}_3}$ members
from ${\bf q}_3=-{\bf q}_1-{\bf q}_2$, viz.,  a contribution from the three-particle correlations and
also a share of the contribution from the
four-particle correlations $\sim \rho_{{\bf q}_1}\rho_{-{\bf q}_1}\rho_{{\bf q}_2}\rho_{-{\bf q}_2}$
(appropriately systematized as regards the primed and the nonprimed
variables (1.11) and (1.12)). Besides, there also arise corrections $\sim \sum_{{\bf q}\neq 0} (\ldots)/N$ to
the zero approximation coefficient functions around $\rho_{\bf q}\rho_{-{\bf q}}$
in (5.14)  (again systematized
as regards the primed and the nonprimed variables). A part of these
corrections can be ``hidden" quite naturally by renormalizing the particle
mass $m$.  After this fixing of the renormalized mass $m^{*}$  we suggest that
$\alpha_q\to 1$  and find the ideal gas density matrix $R_N^0(x|x')$ in which instead of $m$ we
will have $m^{*}$. Then we will address formula (5.16)  and by the same
reasoning we arrive at  (5.1) with the known value of $m^{*}$. Now from
formula (5.16)  we can also answer the question why formula (5.1) holds
true for any  mass $m^{*}$. The matter is that me multiply and divide  by the
same value $R_N^0(x|x')$ ``finding"  the denominator for all the orders of the
perturbation theory by the sums  number and by the wave vector.

    In the approach suggested here the effective mass also arises in the
natural fashion  if we take into consideration the contribution of the
many particle corrrelations  in expressions (1.10) and (2.4) and if we
renormalize the mass by the contributions to be factorized  (possibly with
the dependence upon the wave vector ${\bf q}$)). We will not discuss here higher
approximations and the discussion of the issue of $m^{*}$ will be returned to
in Section VIII.

If we direct  the temperature towards zero,  i.\,e., $\beta\to \infty$ then from (5.1) with
the consideration of the fact that the density matrix  of the ideal
Bose-gas, which in this case is fully degenerate, equals $1/V^N$, we will obtain in
compliance with designation (1.6) the following expression
\ben
R_N(x|x')=e^{-\beta E_0} \psi_0(x')\psi_0(x),
\een
where the normalized wave function of the main state of the interacting
Bose-particles system
\ben
\psi_0(x)={1\over \sqrt {V^N}}\left(\prod_{{\bf q}\neq 0}\sqrt{\alpha_q}
 \right)\exp\left[-{1\over 4} \sum_{{\bf q}\neq 0} (\alpha_q-1)
 \rho_{\bf q}\rho_{-{\bf q}}\right]
\een
coincides  with that discovered for the first time by Bogoliubov and
Zubarev \cite{Zub2}.

    Finally we will consider the classical limit $\hbar \to 0$  of the density
matrix diagonal elements (5.1) when $\rho'_{\bf q}=\rho_{\bf q}$. The energy $E_0$ from (2.16)
in this area boundary transforms into
\be
E_0={N(N-1)\over 2V} \nu_0-\sum_{{\bf q}\neq 0} {N\over 2V} \nu_q.
\ee
Then, the first logarithm  as a sum of ${\bf q}$ in the
exponent  in (5.1) (together with the factor 1/2) is reduced at
$\hbar\to 0$ to ${\rm ln}\ \alpha_q$ and the second one to ${\rm
ln}(1/\alpha_q)$; they  cancel each other, the factor at $\rho_{\bf q}\rho_{-{\bf
q}}$ equaling
\be
&&\nu_q^{*}=\alpha_q\ \tanh \left[{\beta\over 2}E(q)\right]
-\tanh \left[\beta {\hbar^2 q^2\over 4m}\right],\nonumber\\
&&\nu_q^{*}=\beta{N\over V} \nu_q,\ \ \ \hbar\to 0.
\ee
We have made use of the fact that at high
temperatures $m^*\to m$. In the $R_N^0(x|x)$ matrix from (1.9) at $\hbar \to 0$ only an
identical permutation survives
$$R_N^0(x|x)={1\over N!} \left({m\over 2\pi \beta
\hbar^2}\right)^{3N/2}.$$
Consequently, bringing everything together  we have the exact classical
solution for  the density matrix
$$
R_N(x|x)={1\over N!} \left({m\over 2\pi \beta
\hbar^2}\right)^{3N/2} e^{-\beta \Phi},
$$
where the potential energy $\Phi$ is given by  equation  (2.5).

For the partition function(5.13) taking into account the fact that in the
classical limit $\hbar\to 0$ the  structural factor of the ideal gas  (5.8) $S_0(q)=1$ we find
a well-known expression in the random phase approximation \cite{March,Tila}:
\be
Z_N&=&Z_N^0\exp \Bigg\{-\beta {N(N-1)\over 2V}\nu_0\nonumber
\\
&-&{1\over 2} \sum_{{\bf q}\neq 0} \left[
{\rm ln} \left(1+\beta {N\over V}\nu_q\right)-\beta {N\over V}
\nu_q \right]\Bigg\},
\ee
$$Z_N^0={V^N\over N!}\left({m\over 2\pi \beta
\hbar^2}\right)^{3N/2}.$$

Thus from expressions (5,1), (5,13) for the density matrix and the
partition function we have received all the known limiting cases both in the
essentially quantum and classical regions. For that matter we can expect
these expressions to give good results also in the intermediate
temperature region, in particular in the vicinity of the $\lambda$-transition point.

\section{ENERGY}
\setcounter{equation}{0}

From  expression (5.13)  for the partition function we find the
energy by the well-known thermodynamic equation:
$$E={\partial \over \partial \beta} (\beta F)=-{\partial \over
\partial \beta} {\rm ln}\ Z_N.$$
Simple calculations give
\be
E&=&E_0+\sum_{{\bf q}\neq 0} {E(q)\over e^{\beta E(q)}-1}
+\sum_{{\bf q}\neq 0} {\hbar^2 q^2\over 2m^{**}} \left[
{1\over z_0^{-1}e^{\beta\hbar^2 q^2\over 2m^*}-1}-{1\over e^{\beta{\hbar^2
q^2\over 2m^*}}-1}
\right]\nonumber\\
&-&{1\over 2}\sum_{{\bf q}\neq 0} \left({E(q)\over \sinh [\beta
E(q)]}
-{\hbar^2 q^2/2m^{**}\over \sinh [\beta{\hbar^2 q^2/2m^{*}}]}\right)
\nonumber\\
&+&{1\over 2}\sum_{{\bf q}\neq 0}{1\over 1+S_0(q)\left(
\alpha_q\tanh\left[\beta E(q)/ 2\right]-\tanh\left[ \beta
\hbar^2 q^2/ 4m^{*}\right]
\right)}\\
&\times& \Biggl\{ {S_0(q)\over 2} \left[{\alpha_q E(q)\over \cosh^2 \left[\beta
E(q)/2
\right]}-
{\hbar^2 q^2/2m^{**}\over \cosh^2 \left[\beta\hbar^2
q^2/4m^{*}\right]}
\right]
\nonumber\\
&+& {\partial S_0(q)\over \partial\beta}
\left(\alpha_q\tanh\left[{\beta\over 2}E(q)\right]-
\tanh \left[\beta {\hbar^2 q^2\over 4m^{*}}\right]
\right)\Biggr\},\nonumber
\ee

where the value
\be
m^{**}=m^{*}\left/\left[1+ \beta m^{*}{\partial\over \partial
\beta}\left({1\over m^{*}}\right)\right]\right..
\ee

For the actual calculus of the temperature dependence of the
energy $E$ we will write in more detail the ideal Bose-gas
structural factor (5.8) and its derivative with respect to the inverse
temperature
\be
{\partial S_0(q)\over \partial \beta} ={2\over N} \sum_{\bf p}
{\partial n_p \over \partial \beta} n_{|{\bf p}+{\bf q}|},
\ee
which enter (6.1) with the consideration of the Bose--Einstein
condensation phenomenon.

Bose-condensation arises on condition that $z_0=1$ and the critical
temperature $T_c$ is determined from  equation  (5.10) \cite{Huang,Is,Feynm}
\be
T_c={2\pi\hbar^2\over m^{*}} \left[\left.{N\over
V}\right/\zeta(3/2)\right]^{2/3},
\ee
where Riemann $\zeta$-function $\zeta(3/2)=2.612375...$\,. As the effective mass also
happens to be a function of temperature, expression (6.4) as a matter of
fact is the equation for determining the temperature of the Bose-condensation $T_c$.

In (5.4) we will single out the terms with the average number of
particles $n_0$ whose  momenta equal zero:
$$
T\le T_c,\ \ \ z_0=1,
$$
\be
S_0(q)=1+2{n_0\over N} n_q+{1\over N} \mathop{\sum_{{\bf p}\neq
0}}\limits_{{\bf p}+{\bf q}\neq 0} n_p n_{|{\bf p}+{\bf q}|}.
\ee
Now, taking into account the temperature dependence of the
Bose-condensate fraction for an ideal gas \cite{Huang,Is}
\ben
{n_0\over N}=1-\left({T\over T_c}\right)^{3/2},
\een
which is also determined from condition (5.10) for the temperatures $T\le T_c$.
We find from (6.5) that
\ben
S_0(q)={\coth}\left(\beta {\hbar^2 q^2\over 4m^{*}}\right)-
2\left({T\over T_c}\right)^{3/2}n_q
+ {1\over N}\mathop{\sum_{{\bf p}\neq
0}}\limits_{{\bf p}+{\bf q}\neq 0} n_p n_{|{\bf p}+{\bf q}|}.
\een
In this expression we will pass from the summation over the wave
vector  ${\bf p}$  to the integration considering the thermodynamic limit
$N\to \infty$, $V\to \infty$, $N/V=\rho={\rm const}$
and ultimately we find after integration over angles
\be
&&S_0(q)={\coth}\left(\beta {\hbar^2 q^2\over 4m^*}\right)-
{2(T/T_c)^{3/2}\over e^{\beta{\hbar^2 q^2\over 2m^*}}-1}+{m^*\over 4\pi^2 \rho q \hbar^2 \beta}
\nonumber\\
&&\times\int_0^\infty {p\over
e^{\beta{\hbar^2 p^2/2m^*}}-1}{\rm ln} \left|{1-e^{-\beta{\hbar^2(p+q)^2\over 2m^*}}\over
1-e^{-\beta{\hbar^2(p-q)^2\over 2m^*}}}\right|dp,
\ee
$$T\le T_c.$$

For the temperatures which are above the critical temperature the Bose-condensate is absent ($n_0=0$), that is why from (5.8)
\be
S_0(q)&=&1+{m^*\over 4\pi^2 \rho q \hbar^2 \beta}\int_0^\infty {p\over
z_0^{-1}e^{\beta{\hbar^2 p^2/2m^*}}-1}\nonumber\\
&\times&{\rm ln} \left|{1-z_0e^{-\beta{\hbar^2(p+q)^2\over 2m^*}}\over
1-z_0e^{-\beta{\hbar^2(p-q)^2\over 2m^*}}}\right|dp,
\ee

$$ T\ge T_c.$$
We carry out analogous  calculations also for the  structure factor
derivative (6.3):
\be
{\partial S_0(q)\over \partial \beta}&=&3T\left({T\over T_c}\right)^{3/2}{1\over e^{\beta{\hbar^2q^2\over 2m^{*}}}-1}
- {\hbar^2 q^2/4m^{**}\over {\rm
sh}^2\left(\beta{\hbar^2q^2\over 4m^*}\right)}
\left[1-\left({T\over T_c}\right)^{3/2}\right]
\nonumber\\
&-&{m^*\over 16\pi^2 \rho
q \beta m^{**}}\int_0^\infty dp\, {p^3\over \sinh^2\left(
\beta{\hbar^2 p^2\over 4m^*}\right)}\,
{\rm ln} \left|{1-e^{-\beta{\hbar^2(p+q)^2\over 2m^*}}\over
1-e^{-\beta{\hbar^2(p-q)^2\over 2m^*}}}\right|,
\ee

$$T\le T_c$$
and
\be
&&{\partial S_0(q)\over \partial \beta}=
- {m^*\over 16\pi^2 \rho q  \beta m^{**}}
\int_0^\infty dp\,{p^3\over \sinh^2\left[\beta{\hbar^2p^2\over 4m^{*}}-{\ln z_0\over
2}\right]}
\nonumber\\
&&\times\left(1-{\partial \ln z_0\over \partial \beta}\Bigg/{\hbar^2p^2\over 2m^{**}}\right)
{\rm ln} \left|{1-z_0e^{-\beta{\hbar^2(p+q)^2\over 2m^*}}\over
1-z_0e^{-\beta{\hbar^2(p-q)^2\over 2m^*}}}\right|,
\ee
$$T\ge T_c.$$

Let us now analyze  the expression for the energy $E$ from (6.1)  in the
low-temperature region. If $T<T_c$ then $z_0=1$ and the contribution of the
third term  in (6.1) equals zero and the last two terms cancel each other,
and this can   be seen from (6.6) and (6.8) at $\beta\to \infty$,
$S_0(q)\to 1$, $\partial S_0(q)/\partial \beta\to 0$ and also
$\sinh\left[\beta E(q)\right]\sim e^{\beta E(q)}$ as well as
$\cosh^2\left[\beta E(q)/2\right]\sim e^{\beta E(q)}$. Consequently, we obtain Bogoliubov's formula
for the energy as a mean value of the energy of the non-interacting elementary excitations:
$$
E=E_0+\sum_{{\bf q}\neq 0}{E(q)\over e^{\beta E(q)}-1},\ \ \ \ \
T\to 0.$$

In the quasi-classical limit  $\hbar \to 0$ from (6.1) we can easily find:
\be
E={3\over 2} NT+{N(N-1)\over 2V}\nu_0
-{T\over 2} \sum_{{\bf q}\neq 0}{\left(\beta{N\over V}\nu_q\right)^2\over
1+\beta{N\over V}\nu_q}.
\ee
This expression corresponds to the random phases approximation also
for the partition function (5.19). When arriving at it we  took into
account  that at $\hbar\to 0$ the value of $E_0$ in (6.1) gets canceled together with
the second member in the square brackets  of the third term since at high
temperatures the high values of $q$ prove to be important when $\alpha_q\to 1$. The
fourth term in expression (6.1) at $\hbar \to 0$ also  tends to zero and from the
last fifth term we find a contribution  taking into account that  $S_0(q)\to 1$
and $\partial S_0(q)/\partial \beta=0$ as can be seen from (6.7) and (6.9). Finally, the first
term in the square brackets  of the second term of the energy $E$ is the
energy of the ideal gas  (with the temperature dependence of the effective
mass $m^{*}$  being taken into account) and in the classical limit it equals
$3NT/2$.

Hence, both at high and low temperatures  our theory yields correct
limiting cases. That is why we can expect good results also in the
intermediate area of temperatures.

\section{STRUCTURE FACTOR AND THE POTENTIAL ENERGY}
\setcounter{equation}{0}

By definition the structure factor is the mean quadratic fluctuation of
particles density:
\be
S(q)={\int |\rho_{\bf q}|^2 R_N(x|x) dx\over \int R_N (x|x)\,dx}.
\ee
For calculating this expression we will use the  designation of the
partiton function  (3.2) with integration and variables $\rho_{\bf q}$ (5.3) and we will
write (7.1) as a functional derivative:
\be
S(q)=-2{\delta\ {\rm ln}\ Z_N\over \delta \nu_q^{*}},
\ee
where the quantity  $\nu_q^{*}$  is defined  by formula  (5.18). By means of
expression (5.13) we will find from (7.2) that

\be
S(q)={S_0(q)\over
1+S_0(q)\left( \alpha_q\tanh\left[ \beta E(q)/ 2\right]
-\tanh\left[ \beta \hbar^2 q^2/ 4m^{*}\right]\right)}.
\ee

At the temperature of absolute zero ($T=0$~K) when $\nu_q^{*}=(\alpha_q-1)$, we have the expression
which is yielded by Bogoliubov's theory \cite{Bogol,Zub2}:
\be
S(q)={1\over \alpha_q}.
\ee
The classical  limit $\hbar\to 0$ in (7.3) also yields a well-known result for
the structure factor in the random phase approximation:
\be
S(q)={1\over 1+\beta {N\over V} \nu_q}.
\ee
We have one more illustration of the consistency of our theory.

Using formula (7.3) we find the mean value of the potential energy
(2.5) calculated with the full Hamiltonian (1.1):
\ben
\la\Phi\ra_H={\int R(x|x)\Phi\, dx\over \int R(x|x)\,dx}.
\een
Thus,
\be
\la\Phi\ra_H&=&{N(N-1)\over 2V}\nu_0
\\ \nonumber
&+&{N\over 2V}
\sum_{{\bf q}\neq 0}\nu_q\left[
{S_0(q)\over 1+S_0(q)\left(\alpha_q\tanh\left[\beta E(q)/2\right]-\tanh
\left[\beta\hbar^2q^2/4m^{*}\right]\right)}-1\right].
\ee
In its turn, this expression makes it possible to calculate the mean kinetic
energy $K$ as a difference of the full energy (6.1) and (7.5):
\be
K=E-\la\Phi\ra_H.
\ee

\section{EFFECTIVE MASS}
\setcounter{equation}{0}

As we have already mentioned in \cite{JPS97} a different method was
used to obtain a formula for the density matrix  $R_N(x'|x)$ as a product of the
density matrix of the interacting particles with the effective mass
$m^{*}$ multiplied by the factor $P_N(x'|x)$ which takes into account the interparticle
correlations.  Moreover, both for $P_N(x'|x)$ and for $m^{*}$ an expression was found
only in the limit $T=0$~K. The effective mass is determined by the
following equation \cite{JPS97}:
\be
{m\over m^*}=1-{1\over 3N} \sum_{{\bf q}\neq 0}{(\alpha_q-1)^2\over
\alpha_q(\alpha_q+1)}.
\ee
It is curious that this result coincides with the expression for the effective
mass $M^{*}$ of the impurity atom moving  in the liquid $^4$He which we
obtained in \cite{Vak83} using a completely different method
\be
{M\over M^{*}}=1-{1\over 3N} \sum_{{\bf q}\neq 0}{(\alpha_q-1)^2\over
\alpha_q\left(1+\alpha_q {M\over m}\right)}
\left[
1-{\left(1-{M\over m}\right)\over 1+\alpha_q {M\over m}}\right]^2,
\ee
should the atom of the impurity be taken as equal to the mass $m$, i.\,e.,
to the mass of the $^4$He atom. Expression (8.1) with the
square bracket replaced by one was obtained independently in several
studies \cite{Dav,Slyu,Woo,Owen}.

The fact that the ``analytical extension"  of formula (8.2) in the case when
the impurity mass equals that of the  liquid atom takes us to expression
(8.1) testifies to a certain self-consistency of the theory \cite{JPS97}. The same
coincidence of effective masses as shown in \cite{JPS97} cannot be expected at
the temperatures not equaling zero. A dependence between the effective
mass of the impurity $M^{*}$ on the temperature was studied in \cite{VakB}.

It is obvious that for high temperatures the role of interatomic
interaction decreases and $m^*\to m$ when $\beta\to 0$. Let us
consider this question in more detail. Let the impurity atom of
the mass $M$ and  the coordinate ${\bf R}$ move in the liquid. The complete
Hamiltonian of the system ``an atom plus  liquid" consists of the
sum of the Hamiltonians of the liquid (1.1) plus the Hamiltonian of the
impurity
\be
H_i={M{\bf v}^2\over 2}+{N\over V}\bar\nu_0+{\sqrt N\over
V}\sum_{{\bf q}\neq 0}\bar\nu_q e^{i{\bf qR}}\rho_{\bf q}.
\ee
Here the first term is the kinetic
energy of the impurity atom whose velocity equals ${\bf v}$, the two other
terms being its potential energy of the pair interaction with the
liquid atoms with the Fourier $\bar \nu_q$ coefficient. As we consider a
classical system its partition factor $Z$ and aditional energy $\Delta E$ to
the full energy of the liquid (6.10) can be easily found in the
random phase approximation (see (5.19))
\be
Z=Z_NZ_i,
\ee
where
\be
Z_i=V\left({M\over 2\pi \beta \hbar^2}\right)^{3/2}
\exp\Bigg\{-\beta{N\over V}\bar \nu_0+{1\over 2N}\sum_{{\bf q}\neq 0}{\left(\beta{N\over V}\bar\nu_q\right)^2\over 1+\beta{N\over
V}\nu_q}\Bigg\},
\ee
and $Z_N$ can be found from formula (5.19);
\be
\Delta E={3\over 2}T+{N\over V}\bar \nu_0-{T\over 2N}\sum_{{\bf
q}\neq 0}{\left(\beta{N\over V}\bar\nu_q\right)^2\over 1+\beta{N\over V}\nu_q}
-{T\over 2N}\sum_{{\bf
q}\neq 0}\left({\beta{N\over V}\bar\nu_q\over 1+\beta{N\over
V}\nu_q}\right)^2.
\ee

Let us now introduce some effective Hamiltonian of the system ``an atom
plus  liquid" in which the mass of the impurity atom is substituted by the
effective mass $M^{*}$, the interaction between the liquid particles $\nu_q$ is
substituted by $\tilde \nu_q$ at ${\bf q}\neq 0$  and the interaction between the impurity atom
and the liquid is absent.
Constants $\nu_0$ and $\bar\nu_0$ remain invariable
at ${\bf q}=0$ as we are interested by the renormalization via
fluctuation mechanism, when ${\bf q}\neq 0$.
 We will introduce the quantities $M^{*}$, $\tilde \nu_q$ in the
way so that the thermodynamic function of the initial system ``an atom
plus  field" and the approximating model with the effective Hamiltonian
should coincide in the respective approximations. For the suggested
model with the effective Hamiltonian the  partition function can be found
easily
\be
Z&=&Z_N^0V\left({M^{*}\over 2\pi
\beta\hbar^2}\right)^{3/2}\!\!\exp\Bigg[
-\beta{N(N-1)\over 2V}\nu_0-\beta{N\over V}\bar \nu_0\nonumber\\
&+&{1\over 2}\sum_{{\bf q}\neq 0}\beta{N\over
V}\tilde\nu_q-{1\over 2}\sum_{{\bf q}\neq 0}\ln
\left(1+\beta{N\over V}\tilde\nu_q\right)\Bigg]
\ee

Let us now compare expressions for the  partition function of the initial
system(8.4) and (8.5) and the model system (8.7):
\be
&-&{3\over 2}\ln {M\over M^{*}}-{1\over 2}\sum_{{\bf q}\neq 0}
\left[\ln\left(1+\beta{N\over V}\tilde \nu_q\right)-\beta{N\over V}\tilde\nu_q\right]
\nonumber\\
&=&{1\over 2N}\sum_{{\bf q}\neq 0}{\left(\beta{N\over V}\bar\nu_q\right)^2\over 1+\beta{N\over V}\nu_q}
-{1\over 2}\sum_{{\bf q}\neq 0}\left[\ln\left(1+\beta{N\over V}\nu_q\right)-\beta{N\over
V}\nu_q\right].
\ee

Taking into account some arbitrariness in the designation of $M^{*}$ and $\tilde
\nu_q$
we  will equate  in (8.8) logarithmic terms. As a result of this we will
find the  effective potential
\ben
\tilde\nu_q=\nu_q+{\beta\bar\nu_q^2\over V\left(1+\beta{N\over V}\nu_q\right)}
\een
and in the linear approximation the effective mass:
\be
{M\over M^{*}}=1-{1\over 3N}\sum_{{\bf
q}\neq 0}\left({\beta{N\over V}\bar\nu_q\over 1+\beta{N\over
V}\nu_q}\right)^2.
\ee

Let us now put  in (8.9) that  $M=m$, $\bar \nu_q=\nu_q$ and find an expression for the
effective mass $m^{*}$ in the classical limit $\hbar\to 0$:
\be
{m\over m^{*}}=1-{1\over 3N}\sum_{{\bf
q}\neq 0}\left({\beta{N\over V}\nu_q\over 1+\beta{N\over
V}\nu_q}\right)^2.
\ee

Thus, formulae (8.1) and (8.10) determine the effective mass in the
limiting cases $T=0$ and $\hbar\to 0$.
One can rewrite these two formulae using the liquid structure
factor $S(q)$. At the absolute zero temperature from (8.1) and (7.4)
one can find
\be
{m\over m^{*}}=1-{1\over 3N} \sum_{{\bf q}\neq 0}{[S(q)-1]^2\over
S(q)+1},\ \ T=0,
\ee
and from (8.10), (7.5) the high-temperature limit is obtained:
\be
{m\over m^{*}}=1-{1\over 3N} \sum_{{\bf q}\neq 0}[S(q)-1]^2,\ \
T\to\infty.
\ee
The above expressions might be ``sewed together'' numerically and
the temperature dependence of the effective mass can be obtained.
However, both of them are from the source other than the presented
paper. Although, this is necessary at $T=0$ as the mass $m^{*}$ at this condition
just drops out from all the expressions, as seen from (5.1).

To reach full self-consistency of our theory at $T\neq 0$ we need to find the
expression for $m^{*}$ within the considered approach.
One can demand, e.~g., by means of special
selection of $m^{*}$, the expressions for the kinetic energy (7.7)
and that from $\la K\ra=-m\, dF/dm$ (following from the know theorem about
the derivation of the free energy $F$ by some parameter) or $\la K\ra$ calculated
from the density matrix  $R(x'|x)$  from (5.1) to equal.
It is clear that in the exact theory all these expressions
coincide, but as a consequence of the pair correlations
approximation we use here they differ.
To reach the self-consistency of the theory one can utilize the
arbitrariness of the parameter $m^{*}$ and choose it every time to
obtain these formulae to be equal for the kinetic energy at
different temperatures.

A simpler approach to fix $m^{*}$ also exists.
One can, e.~g., choose $m^{*}$ to agree formula (7.3) for
the structure factor with its experimental values at different
temperatures. To make it in average for all the values of the wave vector ${\bf q}$
one can rewrite expression (7.3) as follows:
\ben
{1\over S(q)}-{1\over S_0(q)}-\alpha_q\tanh\left[{\beta E(q)\over 2}\right]
=-\tanh\left(\beta{\hbar^2q^2\over 4m^{*}}\right).
\een
Let us make the integration over ${\bf q}$ in this equation,
adding previously unity in both sides for the sake of convergence.
After the integration over angles we find:
\be
&&\int_0^\infty q^2\left\{{1\over S(q)}-{1\over S_0(q)}-\left(\alpha_q\,\tanh\left[{\beta E(q)\over 2}\right]
-1\right)\right\}\, dq
\nonumber \\
&&=\int_0^\infty q^2\left(1-\tanh\left[\beta{\hbar^2 q^2\over 4m^{*}}\right]\right)\, dq.
\ee
Integral in the right-hand side is easily calculated and we finally
arrive at:
\be
&&\int_0^\infty q^2\left\{{1\over S(q)}-\alpha_q\,\tanh\left[{\beta E(q)\over 2}\right]\right\}\, dq
\nonumber\\
&&=\int_0^\infty q^2\left[{1\over S_0(q)}-1\right]\,dq+\left({m^{*}\over
\beta\hbar^2}\right)^{3/2}\!\!\sqrt{\pi}(\sqrt{2}-1)\zeta\left(3/2\right).
\ee

This equation solves the problem of the effective mass $m^{*}$
calculation via known $S(q)$ for non-zero temperatures.
Let us note that according to (7.4) this quantity equals to
structure factor at $T=0$, and $S_0(q)$ also depends on $m^{*}$. Equation (8.14)
can be used to ``sew together'' formulae (8.11), (8.12) being correct both at low and high temperatures.

Let us indicate one more way to determine $m^{*}$ via the condition
that the pair distribution function of $^4$He at zero interatomic
distance should equal to zero due to hard core in the interaction
potential. Only the numerical calculation can show how all the described approaches to the determination of $m^{*}$
concord.

Therefore, if the liquid structure factor measured experimentally \cite{Ach,Hall,Rob,Sven} is used as a source
information, the theory of liquid $^4$He proposed in this paper
gives the connection between the observable quantities in the
whole temperature interval, including the vicinity of $\lambda$-transition.
Such an approach was utilized by the author in \cite{Vak4,VakHl96} for the
calculation of different physical properties of $^4$He near the absolute zero, $T\to 0$~K.

I express my gratitude to my colleagues from the Department for
Theoretical Physics of the Lviv University, especially to Andrij
Rovenchak, for the discussions about the results.


\def \refname{Reference}

\end{document}